# Initial carrier-envelope phase of few-cycle pulses determined by THz emission from air plasma

Rongjie Xu,[1,2] Ya Bai,[1] Liwei Song,[1] Peng Liu,[1,2,a)] Ruxin Li,[1,2,b)] and Zhizhan Xu[1]

[1]*State Key Laboratory of High Field Laser Physics, Shanghai Institute of Optics and Fine Mechanics, Chinese Academy of Sciences, Shanghai 201800, P. R. China*

[2]*Department of Physics, Tongji University, Shanghai 200092, P. R. China*

The evolution of THz waveform generated in air plasma provides a sensitive probe to the variation of the carrier envelope phase (CEP) of propagating intense few-cycle pulses. Our experimental observation and calculation reveal that the number and positions of the inversion of THz waveform are dependent on the initial CEP, which is near $0.5\pi$ constantly under varied input pulse energies when two inversions of THz waveform in air plasma become one. This provides a method of measuring the initial CEP in an accuracy that is only limited by the stability of the driving few-cycle pulses.

a) Electronic mail: peng@siom.ac.cn

b) Electronic mail: ruxinli@mail.shcnc.ac.cn

The carrier envelope phase (CEP) of ultrashort laser pulses of a few cycles, $\phi$, which is the phase of the carrier oscillations at the instant of maximum amplitude of the pulse envelope, becomes a significant parameter in modifying the electric field of laser pulses and outcomes of the interaction with a medium.[1-3] Intense CEP stabilized few-cycle pulses have been successfully applied in the generation of isolated attosecond pulses and the steering of molecular dissociative ionization dynamics.[4-7] Also, the carrier-envelope phase has profound effects in the weak-field regime, such as on the quantum interference control of semiconductors,[8] and the bound-state atomic coherence.[9]

For fully characterizing the phase of few-cycle laser pulses, several schemes of measuring the actual CEP of few-cycle pulses have been proposed and demonstrated. The stereo-ATI (above-threshold ionization) method measures the ATI electrons in opposite directions along the laser polarization and relates the asymmetry of cut-off electrons to the CEP value.[10,11] The recent advance indicates a capability of single-shot measurement with an accuracy of ~ 100 milli-radian.[12] Another method is the detection of THz emission from air plasma driven by intense few-cycle laser fields, based on the CEP dependent asymmetry of tunneling ionization.[13] Thirdly, the high-harmonic emission at individual half-cycles of a few-cycle pulse has been shown to be able to measure the CEP of the driving laser field *in situ*.[14] These methods all aim to measure the CEP of the localized laser field where intense few-cycle pulses cross with the detection medium by assuming the variation of CEP is negligible during the interaction.

However, for the intense few-cycle laser fields, Gouy phase shift and diffraction dominate the CEP variation even under the condition of linear focusing.[15] For example, if one considers few-cycle laser pulses of 8 *fs* in full width at the half maximum (FWHM) and with the beam diameter of 8 mm, focused by a mirror of $f =$ 200 mm into vacuum, the CEP shift over a distance of 1 mm at the center of focus is 1.43 radian according to the Eq. (21),[16] as shown in Fig. 1(a). Interacting with a nonlinear medium, the variation of CEP becomes more complex because of dispersion of the medium and plasma effect. Taking $N_2$ molecules (with the peak pressure of 100 torr, thickness of 1 mm) ionized by intense few-cycle laser fields (with the pulse energy of 40μJ assuming the diameter of filament is 140 μm), the CEP shift over the interaction length is calculated by solving the propagation equation[17] and plotted in Fig. 1(b). One can see that the CEP variation over the interaction length of 1 mm is 1.34 radian, and the value becomes larger as the focusing length decreases and the pulse energy increases. It is worth noting that the variation curve of the CEP is different from the one in Fig. 1(a).

An alternative option of characterizing the CEP of few-cycle laser pulses is to define the *initial* CEP, which is the actual CEP value of few-cycle pulses before the laser focusing. The initial CEP can be experimentally determined, as will be shown in this work, and its value is free of the uncertainties caused by the focusing and interaction with nonlinear medium. With the knowledge of the initial CEP, the actual CEP of the local laser field interacting with a medium can be estimated in applications.

It is known that the THz waveform reverses its polarity in air plasma because of the variation of the phase and the pulse front (namely the CEP) of the driving few-cycle fields.[18] One can in principal determine the initial CEP of driving few-cycle pulses by taking into account the variation of CEP in the air plasma. However, the value of CEP shift changes as the laser pulse energy varies, so it cannot be pre-determined in experiments, and also, an un-resolved offset of THz signal versus CEP has been reported.[19] Therefore, it is challenging to accurately determinate the initial CEP of few-cycle pulses. In this Letter, we find that the positions where THz waveform reverses are related to the initial CEP of the intense few-cycle fields, from which a method of measuring the initial CEP of few-cycle pulses is given.

In our experiment, we focus on the number and positions of inversions of THz waveforms from different length of air plasma. The CEP stabilized infrared (IR) few-cycle laser pulses [20] (center wavelength of 1.8 μm, 11 fs in FWHM and pulse energy of 0.46 mJ) are focused into the ambient air by a spherical mirror ($f = 150$ mm), and a stable luminescence filament of ∼ 12 mm is formed by the input pulse energy of maximum 450 μJ adjusted by an iris diaphragm. The generated THz waveforms are measured using the balanced diode geometry of electro-optical (EO) sampling method. Inserting a sharp stainless steel blade into the plasma, the THz emission from different lengths of the filament is detected. A sharp piece of quartz wedge is also inserted into the filament at some positions (both in the front and the end of filament) same as the blade and no difference is observed, indicating that no secondary emission from the laser ablation influences the detection of THz emission in the forward direction. Fig. 2(a) shows the THz waveforms as a function of plasma length when the driving few-cycle pulse energy of 400 μJ is used. The inversion of THz emission appears at the position of near 4 mm. By varying the initial CEP stepwise by $0.2\pi$, the THz waveforms are recorded and plotted in Fig. 2(b) - (d), respectively. It is noted that there are two inversions in the evolving THz waveform in Fig. 2(b) and 2(c) but only one in Fig. 2(a) and 2(d). The positions of inversion are moving toward the upstream of air plasma as the initial CEP of the driving pulses increases. When the initial CEP increment changes to $\delta\varphi_0 = 0.6\pi$, the THz waveform inversion position changes back to one as the front inversion moving out of the filament, shown in Fig. 4(d). This observation indicates that the number and positions of the inversions are sensitively determined by the intense CEP of few-cycle laser fields.

THz emissions from air plasma can be described by the integration of far field emission from the transient photocurrents driven by an intense few-cycle laser field propagating in the plasma.[21] The intense few-cycle laser fields in a dispersive medium can be described by the propagation equation[17] in an axial-symmetric coordinates that takes into account diffraction, dispersion, Kerr effect and the polarization caused by photoelectrons from the tunneling ionization of $N_2$ and $O_2$ in air. The collective motion of the tunneling ionized electrons results in a directional nonlinear photocurrent surge [22]

$$\partial_t J_e(\mathbf{r},z,t) + \nu_e J_e(\mathbf{r},z,t) = \frac{e^2}{m}\rho_e(\mathbf{r},z,t)E(\mathbf{r},z,t) \quad , \tag{1}$$

where $\nu_e$, $e$, $m$ and $\rho_e$ denote the electron-ion collision rate, electron charge, mass, and electron density, respectively. The transient current at each propagation step of the calculation is treated as radiation source of far field THz emission [23]

$$E_{THz}(\mathbf{r'},t) = -\frac{1}{4\pi\varepsilon_0}\int \frac{1}{c^2 R}\partial_t J_e(\mathbf{r},z,t_r)d^3\mathbf{r}, \quad (2)$$

The integrated THz amplitude as a function of filament length and initial CEP at different input energy is shown in Fig. 3. It can be seen that the THz amplitude, the number of the THz waveform inversion, and the inversion positions are all related to the initial CEP. As shown in Fig. 3(a), the white color represents that the THz amplitude is zero. Inversions of THz emission appear for some initial CEP of few-cycle pulses at the input energy of 250 μJ. The positions of THz inversion are retrieved for all the initial CEP values in a step size of $0.1\pi$ as plotted in Fig. 3(c). It shows that in the plasma the THz emission crosses with zero only once for most of the initial CEP. For the initial CEP of $0.0\pi$ - $1.0\pi$, the position of inversion moves to the front of the filament and disappears at $\varphi_0 = 0.5\pi$. Increasing the energy of laser pulses to 400 μJ, the starting position of THz generation moves to the upstream of the air plasma because of Kerr effect, as shown in Fig. 3(b) and Fig. 3(d). One can see that for some initial CEP of laser pulses, $\varphi_0 = 0.2\pi$ - $0.4\pi$ in Fig. 3(d), there are two inversions appear in the plasma, and the inversion positions of THz waveform gradually shift to the beginning of the filament. Finally the first inversion disappears at $\varphi_0 = 0.5\pi$. We also find that when the energy of laser pulses increases, the initial CEP of $\varphi_0 = 0.5\pi$, where the first inversion disappears into the beginning of filament, does not change.

The experimental observation is consistent with the calculation result shown in Fig. 3(b) and Fig. 3(d). It is reasonable to consider that the THz modulation by initial CEP of $\varphi_0 = 0.2\pi$ and $0.4\pi$ are corresponding to the conditions of the experimental observation of $\delta\varphi_0 = 0.2\pi$ and $0.4\pi$, as shown in Fig. 2(b) and Fig. 2(c).

The THz emission is intrinsically determined by the asymmetry of induced transient currents in the filament, which is zero at the initial CEP of 0 and maximum at $0.5\pi$ for few-cycle Gaussian pulses.[24] However, there is an intensity-dependent pulse distortion and consequently a CEP shift along the interaction length, because of the Kerr effect and plasma effect.[18] Consequently the relation of THz yield and the corresponding CEP changes, and an offset of the THz signal dependence on the CEP has to be considered.[19] We calculate the offset of the zero THz signal versus CEP in the front part of filament at the input energy of 250 μJ and 400 μJ, respectively, and plot in Fig. 4(a). Inset shows the THz yield as function of CEP at position of A (z = -8.4 mm) of 400 μJ, where we can see that the CEP offset is about $0.5\pi$. The value is similar to that obtained by the input energy of 250 μJ at the position of B (z = -6.0 mm). The offset is due to the asymmetry in the temporal shape of pulse envelope caused by self-steepening because of an intensity-dependent group velocity, which moves the pulse center towards the tail.[25] The THz amplitude is gradually enhanced as function of the plasma length. In Fig. 4(b), it shows the offset trend of the zero THz yield versus CEP at the different energy, and this offset ($0.5\pi$) in the front part of filament is little sensitive to the input energy, which leads to more regular THz inversion in the front part of filament.

The changed trend of the CEP in air plasma, as shown in Fig. 1(b), is one reason for the THz inversion along the filament. In the middle and end part of the filament, the significant increase and decrease of CEP (hump structure) results in the reversed polarity of THz emission, and consequently the integrated THz emission shows an inversion of polarity. As shown in Fig. 3, for the initial CEP of $0.5\pi \leq \varphi_0 < \pi$, the THz polarity is positive at the beginning of the plasma, the superposed THz radiation keeps increasing as the CEP increases in the plasma, since there is no polarity change. It eventually decreases and reverses its polarity at the downstream of the plasma when the increment of CEP is larger than $-\pi$. As a result, an inversion of THz waveform appears in the downstream of plasma. For the initial CEP of $0 \leq \varphi_0 < 0.5\pi$, the THz polarity is negative at the beginning of the plasma, where the CEP can increase to be larger than $0.5\pi$ so that the instantaneous THz emission changes its polarity. As a result, the superposed THz emission becomes zero and reverses its polarity as a function of plasma length. As the energy of laser pulses increases, the variation of CEP in plasma becomes larger than $\pi$, by which the second polarity inversion of THz emission may appear if the first inversion shows at the beginning of the plasma, as shown in Fig. 3(d). Such a variation repeats for the initial CEP of $1.0\pi - 2.0\pi$ except that the polarity of THz waveform reverses.

Such energy-independent CEP values, $\varphi_0 = 0.5\pi$ or $1.5\pi$, at which the inversion of THz waveforms disappears into the front of air plasma, provides a flag that one can use to determine the initial CEP of the driving few-cycle laser pulses.

The accuracy of measuring the initial CEP is determined by how well the transition from two inversions to one is decided. Even through experimentally it is limited by the CEP stability of few-cycle laser pulses, we can calculate the accurate initial CEP value where the inversion disappears at the beginning of filament. By increasing the resolution of the initial CEP to $0.001\pi$ in calculation, we find that the first inversion of THz emission disappears at the initial CEP of $\varphi_0 = 0.500\pi$ - $0.408\pi$ depending on the input energy (200 μJ ~ 900 μJ), as shown in Fig. 4(b). It is a slope with the fitting derivative of $-0.00013$ $\pi \cdot uJ^{-1}$. This variation is due to the self-focusing effect and self-steepening that distorts the envelope of few-cycle pulses in plasma, leading to a little shift off the initial CEP of $0.5\pi$.

There are three regions, I, II and III as shown in Fig. 4(b), for the energy of few-cycle pluses that correspond to different THz inversion phenomena. When the energy is larger than 790 μJ (III), three inversions of THz waveform can be seen for some initial CEP from our calculation result. In this case, the input energy is so high that the pulse envelope is split into two. It is therefore not realistic to discuss CEP related phenomena. For the input energy of 253 μJ - 790 μJ (region II), the number of inversions of THz waveforms along the filament is one or two, corresponding to the cases in Fig. 3(b) and 3(d). The first inversion disappears when the initial CEP is $0.427\pi$ - $0.495\pi$. For the input energy smaller than 253 μJ (region I), the number of inversions of THz waveforms along the filament is one or none for all values of the initial CEP. In this case, the disappearance of THz waveform inversion at the upstream of air plasma indicates the initial CEP is close to $0.5\pi$. It is a good strategy to measure the initial CEP in the experimental condition shown in region I and II. In region II, it is easier to measure the THz waveform from different length of plasma since the pulse energy is higher, but the determined CEP has an error of 73 milli-radian. On the other hand, in region I, the accuracy of initial CEP can be less than 10 milli-radian if one identifies the disappearance of an inversion in the beginning of filament to be $0.5\pi$. Therefore in the practice, the measurement is only limited by the accuracy of the CEP of few-cycle laser pulses.

In conclusion, we investigate the inversion of THz emission in the air plasma driven by the intense few-cycle laser fields. The variation of the number and positions of the inversions are found dependent on the initial CEP of laser pulses. The calculation based on the transient photocurrent model indicates the first inversion vanishing point is $0.5\pi$ in the accuracy that is only limited by the CEP stability of few-cycle laser pulses. For a range of input energies of the driving pulses, the initial CEP by which the inversion disappearing at the beginning of air plasma is little energy-dependent and can be used to determine the initial CEP accurately.

This work is supported by the Chinese Academy of Science, the Chinese Ministry of Science and Technology, and the National Science Foundation of China (Grant Nos. 60978012, 11274326, 11134010 and 11127901), the 973 Program of China (2011CB808103). The first two authors, Rongjie Xu and Ya Bai, make equal contribution to this work.

[1] A. Baltuska, Th. Udem, M. Uiberacker, M. Hentschel, E. Goulielmakis, Ch. Gohle, R. Holzwarth, V. S. Yakovlev, A. Scrinzi, T. W. Hansch, and F. Krausz, Nature **421**, 611 (2003).
[2] A. Apolonski, P. Dombi, G. G. Paulus, M. Kakehata, R. Holzwarth, Th. Udem, Ch. Lemell, K. Torizuka, J. Burgdörfer, T. W. Hänsch, and F. Krausz, Phys. Rev. Lett. **92**, 073902 (2004).
[3] T. Nakajima and S. Watanabe, Phys. Rev. Lett. **96**, 213001 (2006).
[4] M. Hentschel, R. Kienberger, Ch. Spielmann, G. A. Reider, N. Milosevic, T. Brabec, P. Corkum, U. Heinzmann, M. Drescher, and F. Krausz, Nature **414**, 509 (2001); M.F. Krausz and M. Ivanov, Rev. Mod. Phys. **81**, 163 (2009).
[5] F. Ferrari, F. Calegari, M. Lucchini, C. Vozzi, S. Stagira, G. Sansone, and M. Nisoli, Nat. Photon. **4**, 875 (2010).
[6] V. Roudnev, B. D. Esry, and I. Ben-Itzhak, Phys. Rev. Lett. **93**, 163601 (2004).
[7] M.F. Kling, Ch. Siedschlag, A.J. Verhoef, J.I. Khan, M. Schultze, Th. Uphues, Y. Ni, M. Uiberacker, M. Drescher, F. Krausz, and M.J.J. Vrakking, Science **312**, 246 (2006).
[8] T. M. Fortier, P. A. Roos, D. J. Jones, S. T. Cundiff, R. D. R. Bhat, and J. E. Sipe, Phys. Rev. Lett. **92**, 147403 (2004).
[9] Y. Wu and X. X. Yang, Phys. Rev. A **76**, 013832 (2007).

[10]G. G. Paulus, F. Lindner, H. Walther, A. Baltuška, E. Goulielmakis, M. Lezius, and F. Krausz, Phys. Rev. Lett. **91**, 253004 (2003).

[11]T. Wittmann, B. Horvath, W. Helml, M. G. Schatzel, X. Gu, A. L. Cavalieri, G. G. Paulus, and R. Kienberger, Nature Phys. **5**, 357 (2009).

[12]A. M. Sayler, T. Rathje, W. Müller, K. Rühle, R. Kienberger, and G. G. Paulus, Opt. Lett. **36**, 1(2011).

[13]M. Kreß, T. Löffler, M. D. Thomson, R. Dörner, H. Gimpel, K. Zrost, T. Ergler, R. Moshammer, U. Morgner, J. Ullrich, and H. G. Roskos, Nature Phys. **2**, 327 (2006).

[14]C. A. Haworth, L. E. Chipperfield, J. S. Robinson, P. L. Knight, J. P. Marangos and J. W. G. Tisch, Nature Physics **3**, 52 (2007).

[15]M. A. Porras, Phys. Rev. E **65**, 026606 (2002).

[16]M. A. Porras, and P. Dombi, Opt. Express **17**, 19424 (2009).

[17]J. S. Liu, R. X. Li, and Z. Z. Xu, Phys. Rev. A **74**, 043801 (2006).

[18]Y. Bai, L. Song, R. Xu, C. Li, P. Liu, Z. Zeng, Z. Zhang, H. Lu, R. Li, and Z. Xu, Phys. Rev. Lett. **108**, 255004 (2012).

[19]M. D. Thomson, M. Kreß, T. L. Löffler, and H. G. Roskos, Laser Photonics Rev. **1**, 349 (2007).

[20]C. Li, D. Wang, L. Song, J. Liu, P. Liu, C. Xu, Y. Leng, R. Li, and Z. Xu, Opt. Express **19**, 6783 (2011)

[21]K. Y. Kim, J. H. Glownia, A. J. Taylor, and G. Rodriguez, Opt. Express **15**, 4577 (2007); K Y. Kim, A. J. Taylor, J. H. Glownia, and G. Rodriguez, Nature Photon. **2**, 605 (2008).

[22]I. Babushkin, W. Kuehn, C. Köhler, S. Skupin, L. Bergé, K. Reimann, M. Woerner, J. Herrmann, and T. Elsaesser, Phys. Rev. Lett. **105**, 053903 (2010).

[23]C. Köhler, E. Cabrera-Granado, I. Babushkin, L. Bergé, J. Herrmann, and S. Skupin, Opt. Lett. **36**, 3166 (2011).

[24]K. Kim, Phys. Plasmas **16**, 056706 (2009).

[25]Y. Xiao, D. N. Maywar, and G. P. Agrawal, Opt. Lett. **37**, 1271 (2012).

**Figures**:

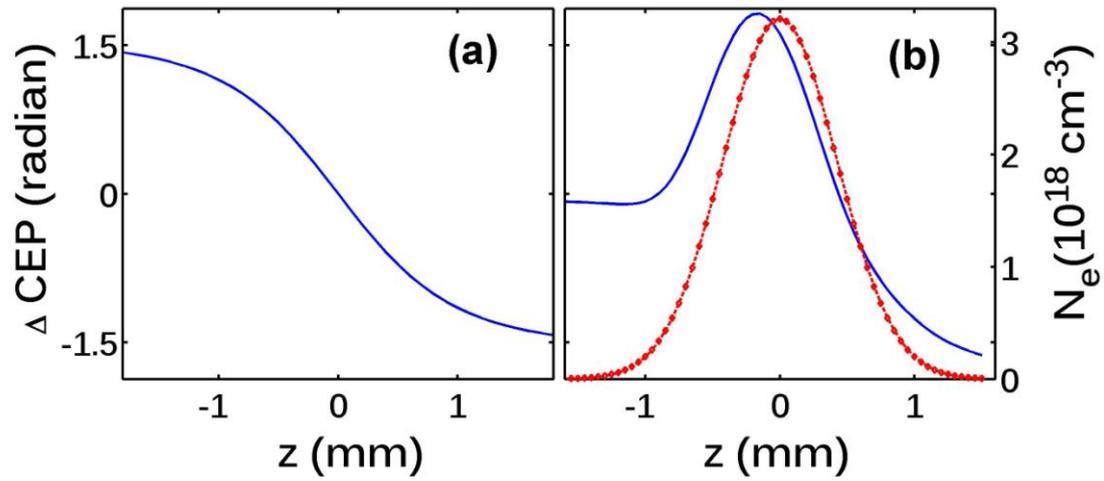

FIG. 1. (a) The CEP shift (blue solid line) around the focus of a few-cycle pulse in vacuum; (b) The CEP shift (blue solid line) in the interaction region between an intense few-cycle laser field and $N_2$ molecules, and the density of molecule number of $N_2$ (red dotted dashed line). See the text for details.

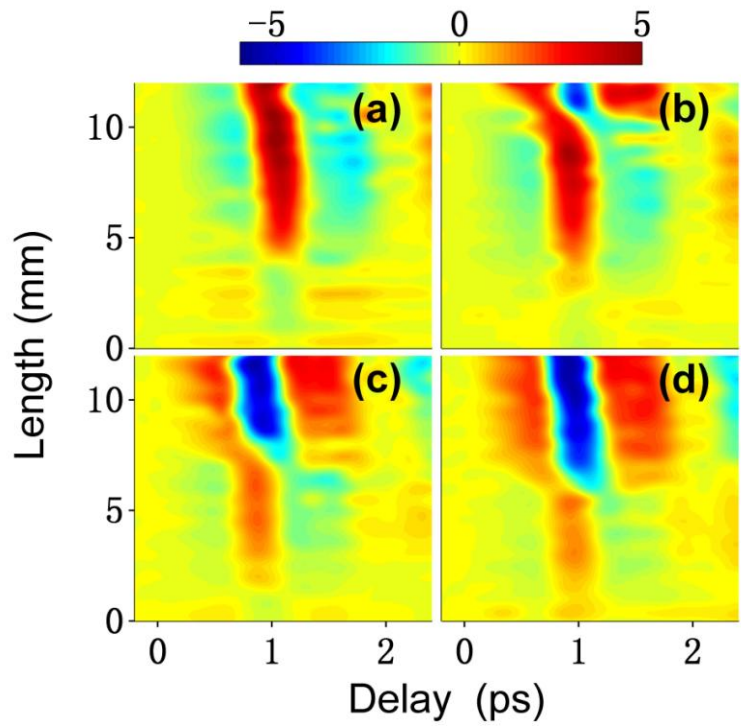

FIG. 2. THz waveform as a function of plasma length by few-cycle laser fields with the initial CEP of (a) - (d): $\delta\varphi_0 = 0.0\pi, 0.2\pi, 0.4\pi, 0.6\pi$, respectively.

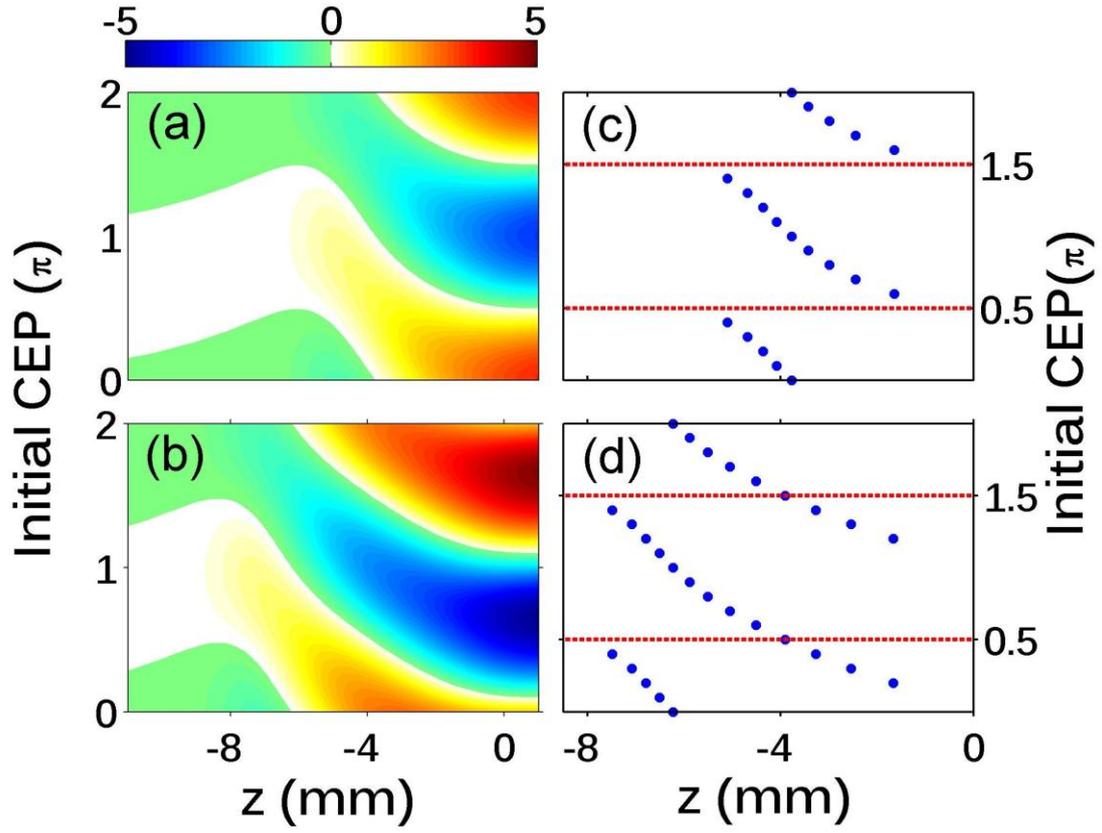

FIG. 3. The calculated THz amplitude as a function of plasma length and the initial CEP of few-cycle pulses with the input energy of (a) 250 μJ and (b) 400 μJ, respectively, and the extracted inversion positions of THz emission in a step-size of 0.1π, for the energy of (c) 250 μJ and (d) 400 μJ, respectively.

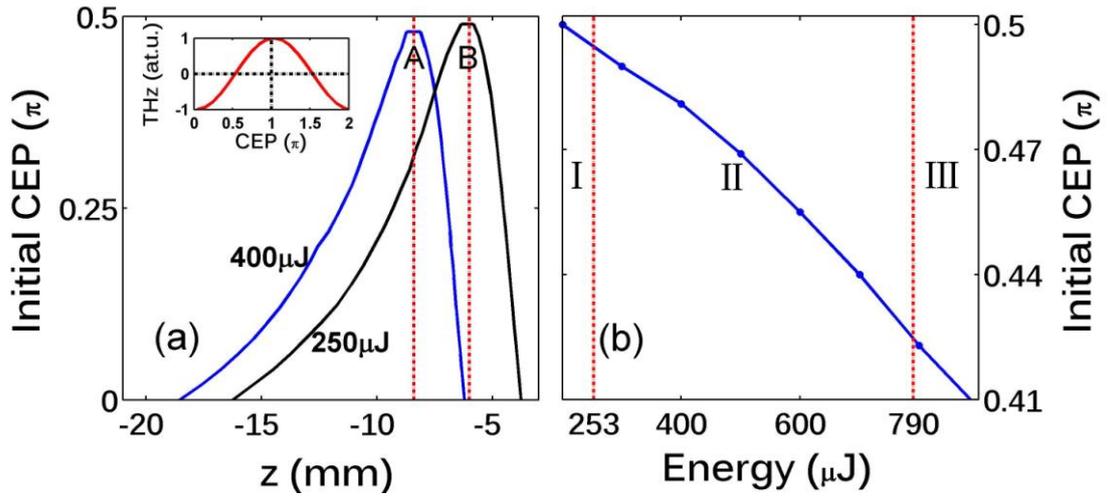

FIG. 4. (a) The calculated offset of zero THz dependence on the CEP along the front part of filament, at the input energy of 250 μJ and 400 μJ, respectively. Inset shows the THz yield as function of CEP at position of A (z= -8.4 mm) of 400 μJ; (b) The calculated initial CEP value by which the inversion of THz emission disappears at the beginning of air plasma under different input energies of 200 μJ ~ 900 μJ.